\documentclass[10pt,letterpaper]{article}
\usepackage[preprint,nonatbib]{customize}
\usepackage[T1]{fontenc}

\usepackage{microtype}
\usepackage[authoryear,round]{natbib}
\setcitestyle{citesep={;},aysep={,},yysep={;}}
\usepackage{graphicx}
\usepackage{booktabs}
\usepackage{float}
\usepackage{placeins}
\usepackage{caption}
\usepackage{needspace}
\usepackage{hyphenat}
\usepackage{iftex}
\ifPDFTeX
\usepackage[utf8]{inputenc}
\fi

\usepackage{amsmath}
\usepackage{amssymb}
\usepackage{enumitem}
\usepackage{url}
\usepackage{tabularx}

\usepackage[hidelinks,pdfusetitle]{hyperref}

\newcommand{\benchname}{JoyFrontBench}
\newcommand{\benchmark}{\textsc{\benchname}}

\newcommand{\qwenmodel}{Qwen3.8 Max (Aug.\ 2026)}
\newcommand{\deepseekmodel}{DeepSeek-V4-Pro-0813}
\newcommand{\qwenmodelcell}[1][l]{\mbox{\qwenmodel{}}}
\newcommand{\deepseekmodelcell}[1][l]{\mbox{\deepseekmodel{}}}

\title{\benchname: Evaluating Front-End Coding Agents through Goal-Oriented Acceptance Contracts}
\author{
  Xiaoyi Qiao\thanks{Equal contribution.} \quad
  Yicheng Wang\footnotemark[1] \quad
  Jianan Lin\footnotemark[1] \quad
  Xiaoyang Ji \quad
  Dongyi Lv \\[0.3em]
  Yidan Huang \quad
  Shi Suo \quad
  Shiying Fan \quad
  Nan Duan\thanks{Corresponding authors.} \quad
  Jiaqi Wang\footnotemark[2] \\[0.5em]
  JD.com
}
\hypersetup{
  pdftitle={\benchname: Evaluating Front-End Coding Agents through Goal-Oriented Acceptance Contracts},
  pdfauthor={Xiaoyi Qiao, Yicheng Wang, Jianan Lin, Xiaoyang Ji, Dongyi Lv, Yidan Huang, Shi Suo, Shiying Fan, Nan Duan, Jiaqi Wang}
}
\date{}
\begin{document}
\raggedbottom
\setlength{\parskip}{4pt}
\maketitle
\vspace{-2em}
\begin{abstract}
Front-end coding agents are expected to produce applications that fulfill user goals, yet high task scores can coexist with failures in essential workflows. Evaluating these agents therefore requires an explicit distinction between partial implementation and application success. We introduce \benchmark{}, a benchmark that makes this distinction through goal-oriented acceptance contracts. Each contract identifies core tasks and mandatory criteria before generation: a core task passes only when all mandatory criteria are satisfied and a shared completion threshold is met, while application success requires every core task to pass on the same operational artifact within a single run. Derived from practitioner requirements, \benchmark{} comprises 100 front-end development problems, 764 tasks, and 2,596 atomic criteria, with reviewed annotations linking user goals to core tasks and evidence requirements. The evaluation framework combines evidence-grounded judgments with deterministic record auditing and fixed scoring rules, preserving graded task assessment alongside explicit core-task and application-level acceptance. We evaluate five coding models under two agent harnesses, OpenHands and OpenCode, on the 70-problem test set; the remaining 30 problems form a public set for validation and diagnostic analyses. Under OpenHands, the mean Average Task Score (ATS) across models is 87.69\%, while the mean Application Success Rate (ASR) is 51.52\%. Further analyses reveal changes in model rankings across harnesses and variation in application success across repeated runs. These findings highlight the limitations of aggregate task scores in characterizing application success and motivate explicit assessment of core-goal fulfillment alongside graded task performance.
\end{abstract}

\section{Introduction}
\label{sec:introduction}

Front-end coding agents translate natural-language requirements into interactive applications. Existing benchmarks assess visual fidelity, functional behavior, and multi-step workflows~\citep{design2code,webgenbench,frontendbench,webcoderbench,webrise,vibecodebench}. These assessments characterize individual aspects of an application; determining whether the application achieves its intended user goal also requires specifying which requirements must hold jointly.

For example, an application may support search, filtering, and editing yet lose saved changes after a reload. Its aggregate task score can remain high even when the persistence failure prevents completion of the primary workflow. Goal-oriented requirements engineering links system requirements to the objectives they serve~\citep{vanlamsweerde2001goals}, while requirements research addresses differences in priority and dependency~\citep{mougouei2020interdependent,svensson2021requirements}. These distinctions motivate an acceptance rule that preserves partial credit while requiring the conditions essential to the user goal.

We introduce \benchmark{}, a front-end application benchmark with \textbf{goal-oriented acceptance contracts}. It comprises 100 problems derived from interviews with more than 30 practitioners, with 764 rubric tasks and 2,596 atomic criteria. Each problem pairs a generation specification with an evaluation rubric linking its primary user workflow to core tasks, mandatory criteria, and validation protocols. Specifications and rubrics undergo three-person review, and reference implementations are checked along their primary workflows. The benchmark is partitioned into a \textbf{70-problem test set} and a \textbf{30-problem public set}.

A core task passes when all mandatory criteria are satisfied and its completion score meets a shared threshold. Application success requires every core task to pass on the same operational artifact. The public set supports threshold selection and diagnostic analyses; the acceptance rule and reporting choices are fixed before test-set evaluation. Evidence-grounded judgments undergo record auditing and deterministic scoring. We rank systems by Application Success Rate (ASR), while retaining Average Task Score (ATS) and Core-Task Pass Rate (CTPR) to characterize partial completion and individual core-task acceptance.

We evaluate five coding models under OpenHands and OpenCode. On the test set under OpenHands, mean ATS, CTPR, and ASR across models are 87.69\%, 80.19\%, and 51.52\%, respectively. High task scores thus coexist with substantially lower application success. Averaged across models, 48.6\% of the 70 test problems yield both successful and unsuccessful outcomes over three OpenHands runs. Model rankings also change across harnesses. These results characterize distinct aspects of the evaluated systems: graded task performance, joint application acceptance, and consistency across repeated runs.

Our contributions are:
\begin{itemize}
    \item \textbf{A benchmark grounded in practitioner requirements.}
    We construct 100 front-end application problems with reviewed specifications and workflow-aligned rubrics that connect user goals to verifiable requirements.
    \item \textbf{Goal-oriented acceptance contracts.}
    We combine mandatory criteria with completion thresholds to define application success on a single operational artifact, alongside graded scores for diagnosis.
    \item \textbf{An empirical characterization of application success.}
    Across five coding models and two harnesses, we examine the relationship between task scores and application acceptance, repeated-run variability, and changes in model rankings across harnesses.
\end{itemize}

\section{Related Work}
\label{sec:related-work}

\paragraph{Front-End Development Benchmarks.}
Design2Code evaluates screenshot-to-code generation by comparing rendered
pages with reference designs~\citep{design2code}. WebUIBench studies
interface reconstruction alongside webpage perception, HTML programming,
and correspondence between interfaces and code~\citep{webuibench}.
FullFront covers webpage design, perception, and code generation.
Its generation tasks include producing code from images or text,
adding interactions, and refining existing code~\citep{fullfront}.

Functional benchmarks examine how generated applications respond to user
actions. WebGen-Bench specifies actions and expected outcomes for
browser-executed tests and assesses website appearance
separately~\citep{webgenbench}. FrontendBench spans tasks ranging from
basic components to complex interactive features and tests their behavior
with predefined interaction scripts~\citep{frontendbench}. Vibe Code Bench
evaluates complete applications built from natural-language specifications.
Its workflows span multiple user actions and include external-service
integrations, allowing evaluation to follow a user request through a
sequence of steps~\citep{vibecodebench}.
\benchmark{} shares this focus on application-level workflows and derives
its problems from requirements elicited through practitioner interviews.
Reviewed specifications and rubrics link the primary user goal to testable
tasks and criteria, making explicit both what the application should
achieve and how it will be assessed.

\paragraph{Evaluation Methods for Front-End Applications.}
Methods for front-end evaluation draw on source code, rendered interfaces,
and runtime interaction. \mbox{ArtifactsBench} combines source code and temporal
screenshots with task-specific checklists to assess the appearance and
dynamic behavior of visual artifacts~\citep{artifactsbench}. WebDevJudge
studies the reliability of automated judges by comparing their judgments
with human preferences over pairs of web implementations. It evaluates
judges under static and interactive settings, with structured rubric
trees guiding human annotation~\citep{webdevjudge}. LiveEvalBench distributes evaluation
across build, code, and browser evaluators, gathering evidence from
deployment, source inspection, and live interaction. Shared rubrics
support comparison across implementations, while implementation-specific
checks accommodate different ways of realizing a
request~\citep{liveevalbench}.

How individual judgments are combined also determines what a benchmark
score represents. \mbox{WebCoderBench} aggregates rule-based and model-based
quality metrics using weights derived from human
preferences~\citep{webcoderbench}. WebRISE maps each requirement to states,
transitions, and assertions. A requirement counts as satisfied only if all
mapped checks pass; the benchmark then reports requirement
coverage~\citep{webrise}. Vibe Code Bench applies a 90\%
substep-success threshold to each workflow and reports the fraction of
passing workflows~\citep{vibecodebench}.

\benchmark{} complements measures of task quality and completion with an
explicit check of whether the application's core goals are met together.
Core tasks and mandatory criteria are fixed before generation; an
application is accepted only if every core task meets all its mandatory
criteria and a shared completion threshold on the same operational
artifact within one run. High scores on other tasks therefore cannot
compensate for a failed core task, while graded task scores still capture
partial completion and support analysis of unmet requirements.

\section{Method}
\label{sec:benchmark-design}

\begin{figure}[H]
    \centering
    \includegraphics[width=\linewidth]{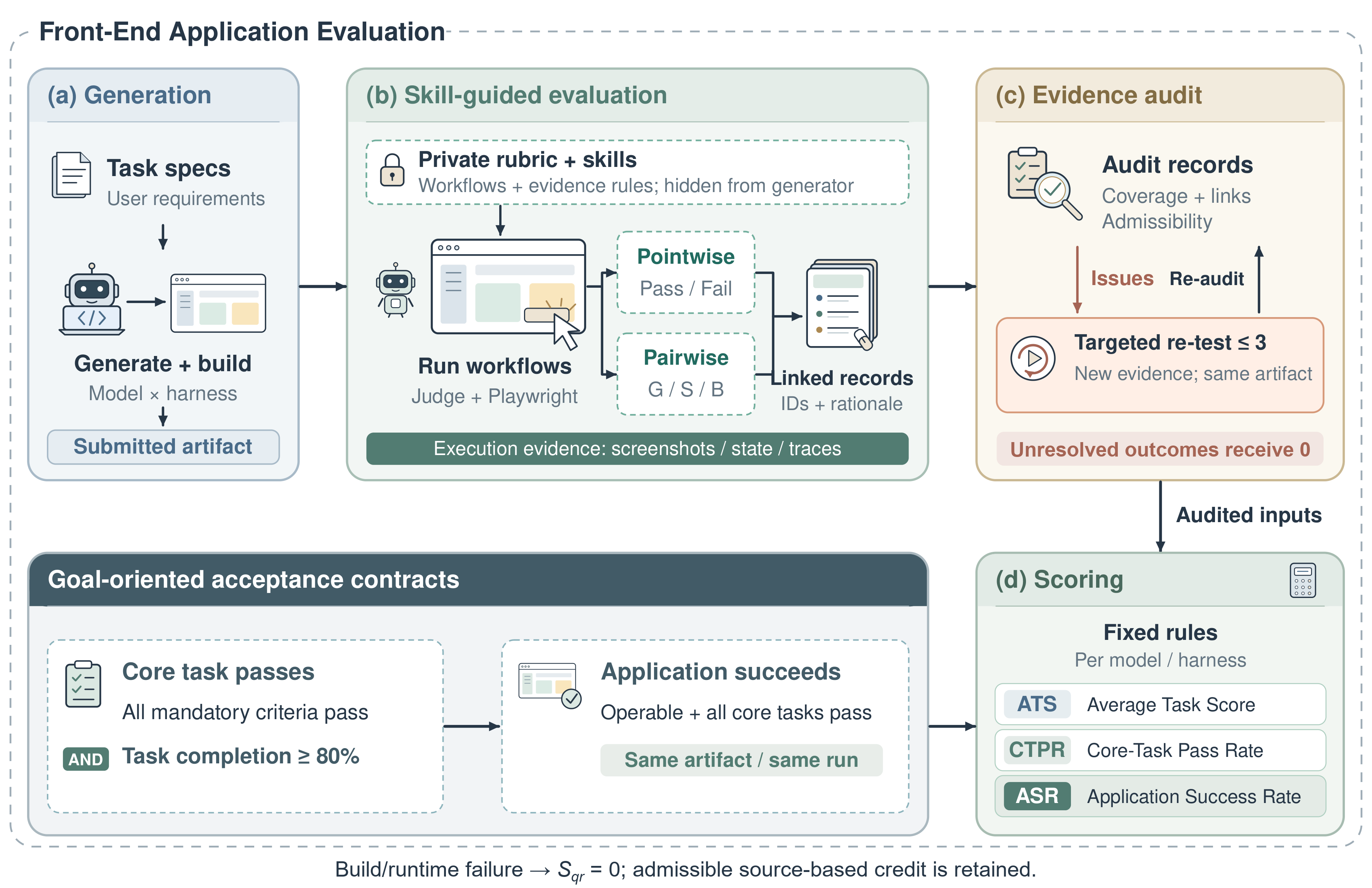}
    \caption{Overview of \benchmark{}. A skill-guided judge records evidence-linked judgments, a deterministic auditor checks record completeness and evidence admissibility, and a model-free scorer computes Average Task Score (ATS), Core-Task Pass Rate (CTPR), and Application Success Rate (ASR). Audit findings trigger up to three targeted remediation rounds before scoring.}
    \label{fig:pipeline}
\end{figure}

\benchmark{} separates graded task performance from application acceptance through \emph{goal-oriented acceptance contracts}. Each contract specifies core tasks, mandatory criteria, and validation requirements. Evaluation collects evidence-linked judgments, audits their records, and applies deterministic scoring rules (Figure~\ref{fig:pipeline}).

\subsection{Benchmark Construction}
\label{sec:acceptance-contract}
\label{sec:data-construction}

\paragraph{Acceptance contracts.}
Annotation starts from the \emph{primary user workflow}: the interactions and resulting states needed to achieve the stated goal, following the goal-oriented view of requirements~\citep{vanlamsweerde2001goals}. Requirements are decomposed into tasks and atomic criteria, each with a validation protocol specifying preconditions, actions, expected observations, and admissible evidence. A criterion is \emph{mandatory} if it verifies the operability of a required step or the correctness of its committed state, including persistence after reload when specified. Pointwise tasks containing mandatory criteria are designated as \emph{core tasks}. Pairwise comparisons against a reference artifact assess subjective presentation quality and contribute to graded task scores; core-task acceptance uses pointwise criteria.

\paragraph{Collection and review.}
We elicit requirements through interviews with more than 30 practitioners in front-end, full-stack, and broader software development, as well as data analysis. Model-assisted rewriting removes organization-specific information and private-service dependencies. Two reviewers independently assess specification clarity, rubric consistency, criterion atomicity, and observability; a third resolves disagreements. Reference implementations are checked along their primary workflows. After resolving review findings, we fix the specifications, rubrics, core-task designations, validation protocols, and operational requirements before generation. Appendix~\ref{app:instance-examples} details construction and annotation. Coding agents generate their own front-end implementations and mock data subject to the requirements stated in each problem specification, including any specified constraints on the data. Evaluation rubrics and reference artifacts are withheld during generation.

\begin{figure}[t]
    \centering
    \includegraphics[width=\linewidth]{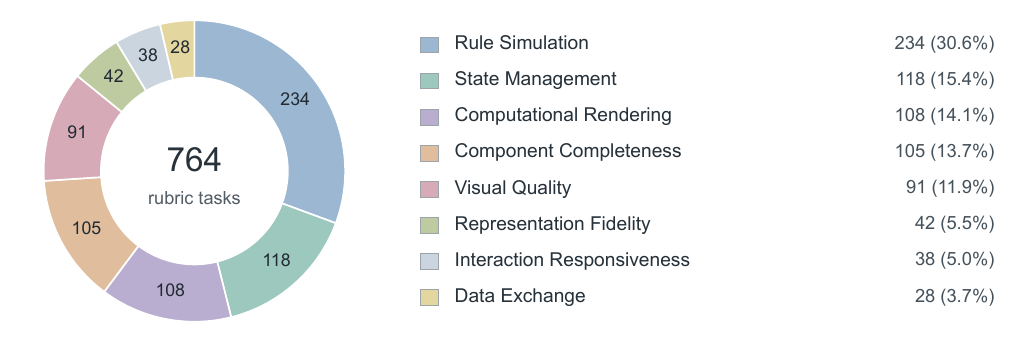}
    \caption{Distribution of the 764 tasks across eight capabilities. Each task has one capability label; definitions appear in Table~\ref{tab:capability-taxonomy}. Sector labels show counts; the legend reports counts and percentages rounded to one decimal place.}
    \label{fig:capability-distribution}
\end{figure}

The benchmark contains 100 problems, 764 tasks, and 2,596 atomic criteria, with 11--52 criteria per problem and one capability label per task (Figure~\ref{fig:capability-distribution}). The 70-problem test set supports the main evaluation. The disjoint 30-problem public set supports selection of the shared completion threshold and reporting rules, together with diagnostic and human-reference analyses. These choices are fixed before test-set evaluation. Specifications, source scenarios, and derived evaluation artifacts are disjoint across the two sets.

\subsection{Evaluation Procedure}
\label{sec:evaluation-pipeline}
\label{sec:generation-harness}

\paragraph{Controlled generation.}
Each run starts in a fresh Docker workspace under OpenHands~\citep{wang2025openhands} or OpenCode~\citep{opencode2026}. Adapters standardize model access, inference settings, dependency and network policies, browser runtime, and outer time and step budgets. Each harness retains its native agent loop, context management, and stopping behavior; the evaluated system is therefore a model--harness pair~\citep{kapoor2025hal,vats2026scaffoldeffect}. Evaluation uses a separate container with the submitted workspace, specification, and scoring policy mounted read-only.

\paragraph{Evidence-grounded judging.}
A GPT-5.5 evaluator with high reasoning effort follows versioned instructions and problem-specific validation protocols. Build and runtime checks provide evidence of operability. Playwright supports interaction recording and collection of screenshots, application state, files, network records, and temporal traces. Behavioral judgments require execution evidence; if launch fails, only criteria predefined as source-verifiable may retain supported credit.

Each pointwise criterion receives a Pass/Fail verdict. Each pairwise task receives one Good/Same/Bad verdict under matched viewports, initial states, interaction paths, and observation times. All verdicts include rationales and evidence links; pairwise judgments require evidence from both artifacts.

\paragraph{Deterministic auditing.}
The auditor checks identifiers, rubric coverage, evidence availability, and evidence-type admissibility. Defects trigger up to three rounds of targeted evidence collection and re-evaluation on the unchanged artifact. Revisions require new evidence, and the full history is retained. Unresolved missing, malformed, or unsupported outcomes receive zero only for the affected criteria or pairwise tasks. The judge supplies criterion verdicts, the auditor checks their supporting records, and a deterministic scorer aggregates the retained results. Appendix~\ref{app:evaluation-schema} gives the record contract and audit workflow.

\subsection{Acceptance Rules and Metrics}
\label{sec:metrics}

\paragraph{Core-task and application acceptance.}
For problem $q$, let $\mathcal{T}_q$ denote its tasks and $\mathcal{C}_q \subseteq \mathcal{T}_q$ its core tasks. A pointwise task $t$ has predefined criteria $\mathcal{J}_{qt}$; a core task additionally has mandatory criteria $\mathcal{M}_{qt} \subseteq \mathcal{J}_{qt}$. In run $r$, let $y_{qrtj}=1$ if a criterion's Pass verdict is retained after auditing and $0$ otherwise. Task completion is
\[
    A_{qrt}
    = \frac{1}{|\mathcal{J}_{qt}|}
      \sum_{j \in \mathcal{J}_{qt}} y_{qrtj},
\]
with all predefined criteria included in the denominator. Using $\tau=0.8$, selected on the public set, acceptance is defined as
\begin{align}
    P_{qrt}
    &= \mathbb{I}[A_{qrt} \geq \tau]
       \prod_{j \in \mathcal{M}_{qt}} y_{qrtj},
    \label{eq:core-task-pass}\\
    S_{qr}
    &= B_{qr}
       \prod_{t \in \mathcal{C}_q} P_{qrt},
    \label{eq:application-success}
\end{align}
where $B_{qr}=1$ only when build and runtime checks pass with a rationale and admissible evidence retained after auditing; otherwise, $B_{qr}=0$.

A core task must satisfy every mandatory criterion and meet the completion threshold. Application acceptance requires operability and acceptance of every core task on the same artifact within a single run. Failed operability sets $S_{qr}=0$, while supported source-verifiable criteria may still contribute to ATS and CTPR.

\paragraph{Scores and aggregation.}
For pointwise tasks, $Z_{qrt}=A_{qrt}$. For pairwise tasks, Good, Same, and Bad map to $Z_{qrt}\in\{1,0.5,0\}$, respectively; unresolved invalid outcomes receive zero. For a fixed model--harness pair, let $\mathcal{Q}$ be the evaluated split and $R=3$ the scheduled runs per problem. Summing over all $q\in\mathcal{Q}$ and $r\in\{1,\ldots,R\}$, we report
\begin{equation}
\begin{aligned}
    \mathrm{ATS}
    &= \frac{1}{|\mathcal{Q}|R}
       \sum_{q,r}
       \frac{1}{|\mathcal{T}_q|}
       \sum_{t \in \mathcal{T}_q} Z_{qrt},\\
    \mathrm{CTPR}
    &= \frac{1}{|\mathcal{Q}|R}
       \sum_{q,r}
       \frac{1}{|\mathcal{C}_q|}
       \sum_{t \in \mathcal{C}_q} P_{qrt},\\
    \mathrm{ASR}
    &= \frac{1}{|\mathcal{Q}|R}
       \sum_{q,r} S_{qr}.
\end{aligned}
\label{eq:metric-aggregation}
\end{equation}

Problems receive equal weight after averaging their scheduled runs. ATS captures pointwise completion and pairwise presentation quality; CTPR measures core-task acceptance; ASR measures application acceptance, including operability. We rank systems by ASR and retain ATS and CTPR for diagnosis. Capability scores average each task over runs, then weight tasks within each capability dimension equally.

\paragraph{Missing and invalid evaluations.}
Infrastructure failures are rerun. Unresolved missing runs or records invalidated as a whole receive zero for all three metrics and remain in the scheduled denominator. Every problem must have a nonempty core-task set for CTPR and ASR to be defined.
Appendix~\ref{app:metric-sensitivity} provides per-run definitions, conditional failure analyses, and threshold sensitivity.

\section{Experiments}
\label{sec:experiments}

\subsection{Experimental Setup}
\label{sec:experimental-setup}

We evaluate five coding models under two agent harnesses,
OpenHands~\citep{wang2025openhands} and OpenCode~\citep{opencode2026}.
OpenHands is the reference harness for the main evaluation; OpenCode supports
cross-harness comparisons. Appendix~\ref{app:harness-config} provides model
details and shared execution controls.

\benchmark{} contains 100 problems: a 70-problem test set and a
disjoint 30-problem public set. The test set supports the main
leaderboard, cross-harness comparisons, repeated-run analysis, and
capability-level evaluation. The public set supports validation and auxiliary
analyses of application failures, representative cases, threshold sensitivity,
and judge--human agreement. We schedule three independent generation-and-evaluation runs per
model--harness--problem condition, with a five-hour generation budget and up
to 500 harness-native steps or iterations per run. Evaluation allows up to
three targeted remediation rounds. Scoring and aggregation are detailed in
Appendix~\ref{app:metric-sensitivity}.

\subsection{Overall Benchmark Performance}
\label{sec:overall-performance}

Table~\ref{tab:main_leaderboard} reports task scores, core-task acceptance, and
application success, together with harness comparisons and repeated-run outcomes.

\begin{table}[H]
\centering
\small
\captionsetup{position=bottom}
\setlength{\tabcolsep}{3pt}
\begin{tabular*}{\linewidth}{@{\extracolsep{\fill}}lrrrrrrrr@{}}
\toprule
& \multicolumn{3}{c}{Performance (OH)}
& \multicolumn{2}{c}{Harness Comparison}
& \multicolumn{3}{c}{Repeatability (OH)} \\
\cmidrule(lr){2-4}\cmidrule(lr){5-6}\cmidrule(l){7-9}
Model & ATS & \textsc{CTPR} & \textsc{ASR} & OC \textsc{ASR}
& $\Delta$ \textsc{ASR} & All pass & Mixed & All fail \\
\midrule
\qwenmodelcell{}     & 89.56 & \textbf{83.02} & \textbf{57.62} & \textbf{56.67} & +0.95 & 38.6 & 37.1 & 24.3 \\
Kimi K3         & 88.98 & 82.62 & 56.67 & 45.71 & +10.95 & 20.0 & 65.7 & 14.3 \\
GPT-5.6 Sol     & \textbf{90.53} & 82.34 & 55.71 & 46.67 & +9.05 & \textbf{40.0} & 37.1 & 22.9 \\
\deepseekmodelcell{} & 85.77 & 80.16 & 47.62 & 35.71 & +11.90 & 21.4 & 51.4 & 27.1 \\
GLM-5.3         & 83.63 & 72.82 & 40.00 & 49.05 & $-9.05$ & 14.3 & 51.4 & 34.3 \\
\midrule
Mean            & 87.69 & 80.19 & 51.52 & 46.76 & +4.76 & 26.9 & 48.6 & 24.6 \\
\bottomrule
\end{tabular*}
\caption{Main evaluation results on the 70-problem test set, ordered by
OpenHands (OH) \textsc{ASR}. Performance and harness scores (\%) average three
scheduled runs per problem, then macro-average across problems
($\tau=0.8$). $\Delta$ \textsc{ASR} is OH minus OpenCode (OC) in percentage
points, computed before rounding. OH repeatability (\%) uses 70 model--problem
conditions per model, each with three scored runs (210 runs per model). The mean row averages five
models. Bold marks the highest performance scores and all-pass share.}
\label{tab:main_leaderboard}
\end{table}

\paragraph{Overall performance.}
On the test set, the five models achieve a mean ATS of 87.69\% and a mean
\textsc{ASR} of 51.52\% under OpenHands. High task scores thus coexist with
substantially lower application success. GPT-5.6 Sol
achieves the highest ATS (90.53\%) but ranks third in \textsc{ASR} (55.71\%),
while \qwenmodel{} has the highest \textsc{ASR} (57.62\%). The top three
\textsc{ASR} point estimates differ by less than two points. \textsc{CTPR}
measures acceptance of individual core tasks, complementing the joint
application acceptance measured by \textsc{ASR} (Section~\ref{sec:metrics}).

\paragraph{Harness sensitivity.}
Test-set performance varies across agent harnesses. The OpenHands--OpenCode
\textsc{ASR} difference ranges from $-9.05$ to $+11.90$ percentage points.
GLM-5.3 scores 49.05\% under OpenCode versus 40.00\% under OpenHands, ranking
second under OpenCode but fifth under OpenHands. Kimi K3 and \deepseekmodel{}
instead score 10.95 and 11.90 points higher under OpenHands, respectively.
These results characterize the complete model--harness configuration.

\paragraph{Run-to-run variability.}
Each model has three scored OpenHands runs for each of the 70 test problems
(210 runs per model). The share of model--problem conditions with both
successful and unsuccessful outcomes averages 48.6\% across models. Kimi K3 has the
highest mixed-outcome share (65.7\%), while GPT-5.6 Sol has the highest
all-success share (40.0\%). Similar aggregate \textsc{ASR} values can thus
accompany different repeated-run outcomes; these runs include both generation
and evaluation (Appendix~\ref{app:reliability}).

\begin{figure}[H]
\centering
\includegraphics[width=\linewidth]{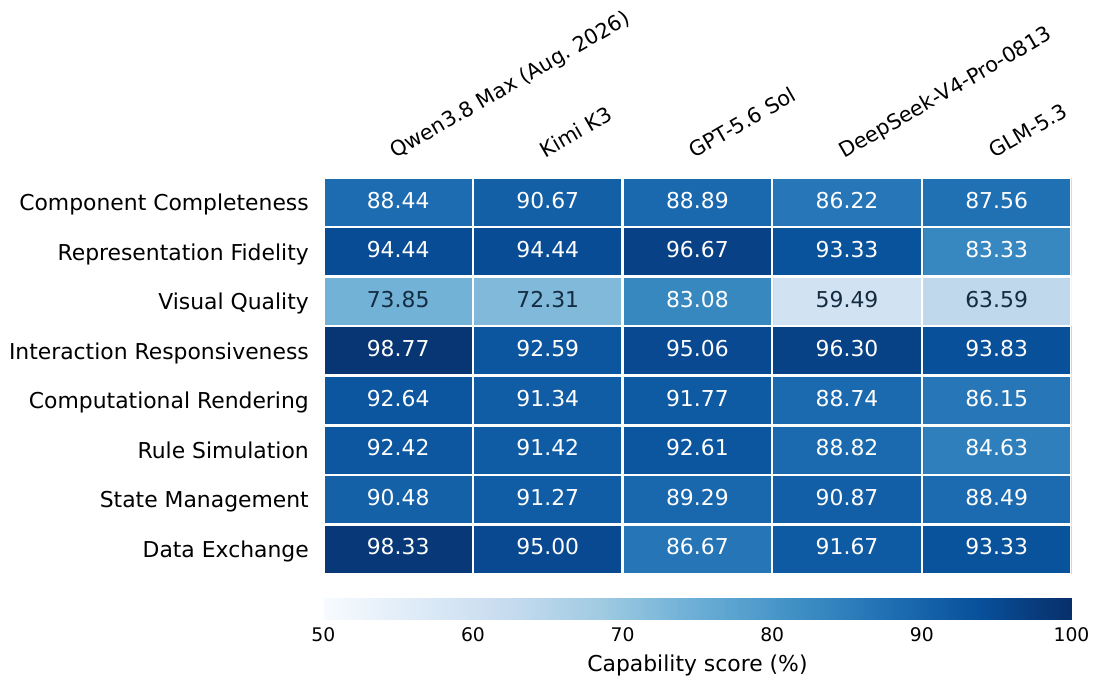}
\caption{Capability scores (\%) on the 70-problem test set under
OpenHands. Cell labels give mean task scores; colors encode
absolute values on a 50--100 scale. Full scores and task counts are in
Table~\ref{tab:capability_scores} (Appendix~\ref{app:capability-results}).}
\label{fig:capability_scores_heatmap}
\end{figure}

\paragraph{Capability-wise analysis.}
Figure~\ref{fig:capability_scores_heatmap} reports scores across eight
capabilities on the test set under OpenHands. Visual Quality is the lowest-scoring
dimension for every model (cross-model mean 70.46\%), compared with 95.31\%
for Interaction Responsiveness. GPT-5.6 Sol achieves the highest Visual
Quality score (83.08\%). These task-level diagnostics complement
application-level success; category-level \textsc{ASR} is in
Appendix~\ref{app:category-asr}.

\subsection{Understanding Application Failures}
\label{sec:application-failures}

On the public set, 32.3\% of unsuccessful OpenHands runs with valid criterion
judgments achieve at least 90\% ATS. These runs show how high task scores can
coexist with failures in the core workflow.

\paragraph{Failure mechanisms.}
Table~\ref{tab:failure_mechanisms} characterizes mandatory-criterion and
completion failures among all unsuccessful OpenHands runs on the public set
with valid criterion judgments. Mandatory-only failures
account for 54.4\% of valid failed runs, compared with 28.6\% completion-only
and 17.1\% combined failures. The mandatory-only share ranges from 43.2\% for
\qwenmodel{} to 63.0\% for GLM-5.3.

\begin{table}[H]
\centering
\small
\captionsetup{position=bottom}
\begin{tabular*}{\linewidth}{@{\extracolsep{\fill}}lrrr@{}}
\toprule
Model & Mandatory only & Completion only & Both \\
\midrule
\qwenmodelcell{}     & 43.2\% & 37.8\% & 18.9\% \\
Kimi K3         & 53.8\% & 33.3\% & 12.8\% \\
GPT-5.6 Sol     & 52.5\% & 22.5\% & 25.0\% \\
\deepseekmodelcell{} & 55.3\% & 29.8\% & 14.9\% \\
GLM-5.3         & 63.0\% & 22.2\% & 14.8\% \\
\midrule
All models      & 54.4\% & 28.6\% & 17.1\% \\
\bottomrule
\end{tabular*}
\caption{Failure composition among unsuccessful OpenHands
public-set runs with valid criterion judgments (\%). ``Both''
indicates a mandatory-criterion failure in one core task and a completion
failure in another core task whose mandatory criteria pass.
The final row pools valid failed runs across models; rows may differ from
100\% by rounding. The three categories cover the valid failed runs analyzed.}
\label{tab:failure_mechanisms}
\end{table}

Mandatory-criterion failures arise when a required workflow step or
committed-state condition is not satisfied, for example, when an application
accepts invalid input or fails to synchronize dependent views.
A completion failure occurs when a core task satisfies its mandatory criteria
but falls below the completion threshold. These mechanisms correspond to the
two conditions in the core-task acceptance rule.

\paragraph{Representative cases.}
Two public-set artifacts illustrate these mechanisms. In Q20, a
financial dashboard achieves 90.48\% ATS but fails critical transfer validation
and state consistency: a negative input is silently converted to a positive amount and reaches confirmation, while a valid
transfer does not update the budget view. Failed mandatory criteria give
$S_{qr}=0$ despite high graded completion.

\begin{figure}[H]
\centering
\begin{minipage}[b]{0.48\linewidth}
\centering
\includegraphics[width=\linewidth,trim=0 175 0 614,clip]{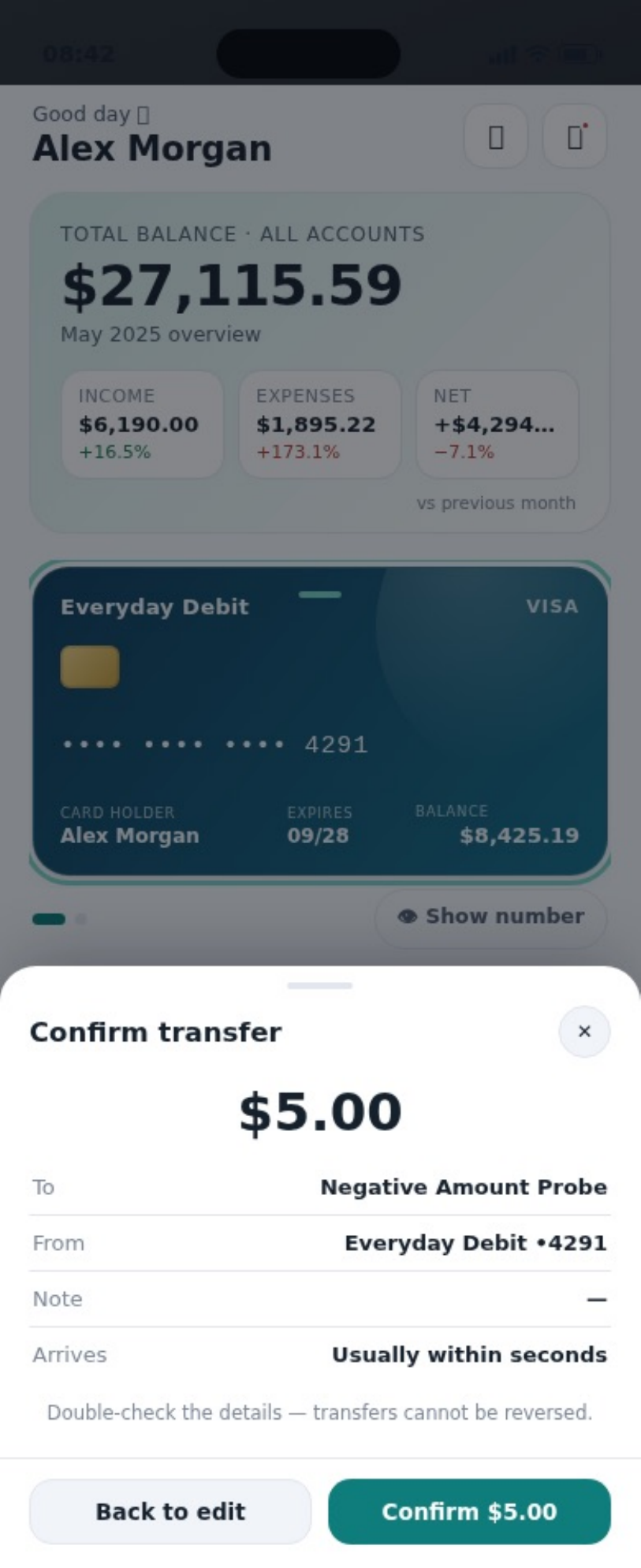}\\
{\footnotesize (a) Q20: failed transfer validation}
\end{minipage}\hspace{0.03\linewidth}
\begin{minipage}[b]{0.47\linewidth}
\centering
\includegraphics[width=\linewidth]{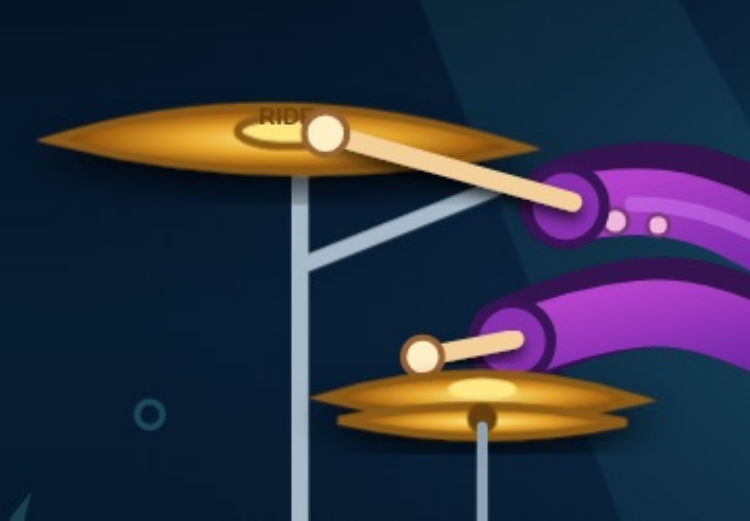}\\
{\footnotesize (b) Q29: unmet completion rule}
\end{minipage}
\caption{Audited public-set cases with $S_{qr}=0$.
(a) Q20: an invalid negative input leads to a \$5.00 confirmation (failed mandatory criterion).
(b) Q29: ride-cymbal contact frame; lack of synchronized response is established
by runtime samples in Appendix~\ref{app:metric-case-evidence} (incomplete core task).}
\label{fig:representative-failures}
\end{figure}

In Q29, a Scalable Vector Graphics (SVG) animation satisfies all six mandatory criteria, but two core tasks
score only 75\% and 50\%, below the 80\% completion threshold. The arms lack
natural taper, and struck drum surfaces do not respond in synchrony with the
sticks. The application thus has $S_{qr}=0$ despite 87.50\% ATS. Q20
exposes critical business-rule and cross-view state failures, whereas Q29
exposes incomplete core-task behavior. Table~\ref{tab:metric-case-summary}
contrasts their per-run scores, and Figure~\ref{fig:representative-failures}
shows the artifacts. Appendix~\ref{app:metric-case-evidence} provides detailed
evidence; Appendix~\ref{app:end-to-end-cases} gives additional cases.

\begin{table}[H]
\centering
\small
\caption{Audited single-run results from the public set. ATS, core-task pass
fraction, and mandatory-criterion pass rate are percentages;
$S_{qr}$ is the binary application-success indicator.}
\label{tab:metric-case-summary}
\begin{tabular*}{\linewidth}{@{\extracolsep{\fill}}llrrrr@{}}
\toprule
Case & Artifact & ATS & Core pass (\%) & $S_{qr}$ & Mandatory pass (\%) \\
\midrule
A & Q20 financial dashboard & 90.48 & 50.00 & 0 & 75.00 \\
B & Q29 octopus drummer & 87.50 & 33.33 & 0 & 100.00 \\
\bottomrule
\end{tabular*}
\end{table}

\FloatBarrier

\subsection{Evaluation Reliability and Threshold Sensitivity}
\label{sec:human-alignment}

Given reported disagreement between model-based web evaluators and human
preferences~\citep{webdevjudge}, we compare automated judgments with an
adjudicated human reference on the 30-problem public set. Two evaluators independently
annotate the artifacts under a shared protocol, with a third adjudicating
disagreements. The automated judge has no access to these human labels.
The study compares 726 pointwise criteria and 15 pairwise tasks
(Appendix~\ref{app:human-materials}). The pointwise comparison yields
83.47\% exact agreement, Cohen's $\kappa=0.651$, and Macro-F1 of 0.825. For problem-level
scores aggregated from pointwise judgments, Pearson $r=0.807$ and Spearman
$\rho=0.766$. The root-mean-square error normalized by the standard deviation
of the human scores (RMSE/SD) is 0.631, and mean error is $+0.0362$.

Label agreement measures consistency of criterion verdicts; correlations
measure association between scores, while RMSE/SD and mean error quantify
their differences. Appendix~\ref{app:human-materials} gives the full results
and metric definitions.

\paragraph{Threshold sensitivity.}
On the public set under OpenHands, mean \textsc{ASR} falls from 58.22\% at $\tau=0.75$ to 50.89\% at
$\tau=0.80$ and 44.89\% at $\tau=0.90$. At $\tau=0.80$, the finite number of
criteria means that 86.6\% of core tasks require every criterion to
pass. The threshold determines how many non-mandatory criteria may fail;
every mandatory criterion must pass at all thresholds. We use $\tau=0.80$ for the
primary acceptance rule; full sensitivity results are in
Appendix~\ref{app:threshold-sensitivity}.

\FloatBarrier

\section{Conclusion}
\label{sec:conclusion}

\benchmark{} evaluates application success through goal-oriented acceptance contracts. Mandatory criteria and completion thresholds define core-task acceptance, and all core tasks must pass on the same operational artifact. The resulting metrics connect graded task performance, individual core-task acceptance, and joint application success.

Experiments on the 70-problem test set show that high task scores coexist with substantially lower application success, while repeated runs and different harnesses yield variation in outcomes and model rankings. Analyses on the 30-problem public set identify mandatory-rule violations, inconsistent committed states, and incomplete core behaviors in unsuccessful applications. These findings support evaluating core-goal fulfillment alongside partial completion, with results interpreted under the reported execution conditions.

\paragraph{Limitations.}
\label{sec:limitations}
The benchmark draws on one organizational setting and uses Chinese-language specifications for self-contained front-end applications. Broader coverage of organizations, languages, and applications with external service dependencies would strengthen its generality. Acceptance is determined by the reviewed rubrics and the completion threshold selected on the public set. Human-reference evaluation covers atomic verdicts and pointwise scores; direct validation of application-level acceptance remains an open task. The three repeated runs capture variation in both generation and evaluation. The reported rankings describe the tested model--harness configurations and budgets.

\paragraph{Use of AI tools.}
AI tools assisted with language editing and consistency checks during manuscript preparation. Their use in benchmark construction and evaluation is described in Section~\ref{sec:benchmark-design} and Appendices~\ref{app:instance-examples} and~\ref{app:evaluation-schema}. The authors retain responsibility for the manuscript's content.

\label{page:main-end}
\par\bigskip
\bibliographystyle{plainnat}
\begingroup
\setlength{\bibsep}{2pt}
\bibliography{refs}
\endgroup
\clearpage
\appendix
\section{Benchmark Construction and Rubric}
\label{app:instance-examples}

\subsection{Construction and Human Review}

The benchmark contains 100 distinct Chinese-language problem specifications,
with 764 tasks and 2,596 atomic criteria (11--52 criteria per problem).
Specifications and evaluation artifacts were constructed in four stages:
\begin{enumerate}[leftmargin=*]
    \item \textbf{Needs elicitation.} In-depth interviews with more than 30
    front-end, full-stack, and software engineers and data analysts at JD.com
    elicited front-end generation cases, product requirements, and full-stack
    development requests as candidate scenarios.
    \item \textbf{De-identification and reformulation.} A structured-output
    GPT-5.6 Sol pipeline removed proprietary identifiers and sensitive business
    information, extracted the front-end objective, and eliminated dependencies
    on databases, back-end services, internal APIs, and organization-specific
    infrastructure. The evaluated model is required to generate the front-end implementation and mock data
    according to the specification, including any stated constraints on the data.
    \item \textbf{Human review.} Three reviewers with at least one year of
    practical front-end development experience followed a shared protocol.
    Two independently checked that every scored requirement followed
    unambiguously from the specification for a qualified engineer, terminology
    and state rules were consistent, no criterion combined independently
    falsifiable requirements, and every validation step could yield observable
    evidence. The third reviewer adjudicated disagreements.
    \item \textbf{Implementation-based refinement.} Each reviewed
    specification was implemented by a frontier coding model and exercised by
    hand along its primary user workflow. Specification gaps were revised before
    regenerating the artifact; implementation errors were repaired. Iteration
    continued until reviewers accepted the core functions, interaction loops,
    state rules, and visual requirements. The reviewed specifications,
    tasks, and criteria were then frozen. The resulting reference artifact is the
    pairwise Good/Same/Bad comparison target.
\end{enumerate}

\subsection{Rubric Annotation and Capability Taxonomy}
\label{app:capability-taxonomy}

The generation specification defines the artifact, functional and interaction
requirements, visual and layout requirements, and engineering constraints.
The evaluation rubric additionally contains capability labels,
tasks, criteria, and validation protocols. A task is a
semantically coherent and independently evaluable objective; its criteria are
atomic checks with explicit preconditions, actions, observations, and expected
outcomes. Mandatory criteria identify operability or committed-state
requirements along the primary user workflow. Tasks containing mandatory
criteria are designated as core tasks; pairwise presentation tasks have no
mandatory criteria.

The complete benchmark has five mutually exclusive application categories:
Web and Business Interfaces (30 problems), Demonstrations and Concept
Visualizations (23), Applications (21), SVG and Vector-Graphics Generation
(13), and Data Visualization (13). At task level, exactly one capability
label is assigned to each task, so capability counts do not overlap.

Table~\ref{tab:capability-taxonomy} provides operational definitions and the annotation distribution.

\begin{table}[htbp]
    \centering
    \small
    \caption{Operational definitions and annotation distribution of the eight capability dimensions. Percentages are calculated from unrounded counts and displayed to one decimal place; consequently, they sum to 99.9\%.}
    \label{tab:capability-taxonomy}
    \begin{tabularx}{\linewidth}{>{\raggedright\arraybackslash}p{0.22\linewidth}Xrr}
        \toprule
        Capability & Operational definition & Tasks & Share \\
        \midrule
        Component Completeness & Completeness of required pages, components, controls, information modules, and necessary technical elements & 105 & 13.7\% \\
        Representation Fidelity & Fidelity with which a static view represents specified entities, data, structures, or domain objects & 42 & 5.5\% \\
        Visual Quality & Coherence of composition, hierarchy, typography, color, and overall style for the target setting & 91 & 11.9\% \\
        Interaction Responsiveness & Whether direct actions such as clicks, drags, scrolling, and keyboard input produce the expected response & 38 & 5.0\% \\
        Computational Rendering & Whether changes in inputs or parameters correctly and synchronously update derived values, content, and visual output & 108 & 14.1\% \\
        Rule Simulation & Correctness of implemented business rules, physical mechanisms, state machines, or deterministic algorithms & 234 & 30.6\% \\
        State Management & Consistency of state across multi-step operations, views, undo/redo actions, or page reloads & 118 & 15.4\% \\
        Data Exchange & Correctness of content, structure, and format during data import, export, or transfer across system boundaries & 28 & 3.7\% \\
        \midrule
        \textbf{Total} & --- & \textbf{764} & \textbf{100\%} \\
        \bottomrule
    \end{tabularx}
\end{table}

\subsection{Benchmark Materials}

The benchmark links each specification to its tasks, criteria, judging modes,
capability labels, and mandatory-criterion designations. The scorer binds
core-task policies to versioned specifications and rejects mismatched rubrics.

\Needspace{12\baselineskip}
\section{Harness Controls and Execution Budgets}
\label{app:harness-config}

\paragraph{Evaluated systems.}
The five evaluated models are \qwenmodel{}~\citep{alibaba2026qwen38max},
Kimi K3~\citep{moonshot2026kimik3}, GPT-5.6 Sol~\citep{openai2026gpt56solmodel},
\deepseekmodel{}~\citep{deepseek2026v4pro}, and
GLM-5.3~\citep{zai2026glm53}. \qwenmodel{} denotes the August 2026 general-availability release; \deepseekmodel{} denotes the August 13, 2026 release.
Provider documentation lists image input for the first three
\citep{alibaba2026qwen38max,moonshot2026kimik3,openai2026gpt56solmodel}
and text-only input for GLM-5.3~\citep{zai2026glm53model}.

Two benchmark-specific adapters connect the unchanged upstream harnesses to
a shared model registry. They standardize model access, inference
configuration, workspace initialization, and benchmark execution. Each
model--harness--problem--run combination starts independently. OpenHands is the
reference harness for the main leaderboard and failure analysis; OpenCode
supports cross-harness comparisons.

\paragraph{Controlled settings.}
Both harnesses receive the same benchmark specification, clean Docker
workspace, network and dependency policies, Playwright 1.58.1 runtime, shared
model registry, and five-hour wall-clock budget. We additionally impose a
nominal cap of 500 iterations in OpenHands and 500 steps in OpenCode.
Iteration and step counts use each harness's native execution units.

Each harness retains its native agent loop, context-management strategy,
tool/action representation, execution behavior, and stopping logic. These
components remain part of each evaluated model--harness system.

\section{Evaluation Skills, Report Contract, and Evidence Audit}
\label{app:evaluation-schema}

Each artifact is evaluated in an isolated Docker container with the submitted
workspace, specification, and scoring policy mounted read-only. Validation
first checks the artifact contract and attempts the required build and launch.
A build or runtime failure sets $B_{qr}=0$. If the artifact cannot launch,
only criteria predefined as source-verifiable may receive supported credit.
The judge is GPT-5.5 with high reasoning effort.

The automated judge follows the benchmark-wide pipeline skill, the common
web-evaluation skill, and the evaluation profile for the current problem. Each
profile specifies the required workflows, atomic criteria, and admissible
evidence modalities.

For pointwise tasks, the evaluation record associates each atomic criterion
with its own Pass/Fail verdict, rationale, and supporting evidence.
For pairwise tasks, it retains one task-level Good, Same,
or Bad verdict, evidence from both the candidate and reference artifacts, and
the viewport, initial state, interaction path, and observation time used for
comparison. The deterministic report writer matches these results to the
authoritative task and criterion definitions. Missing criteria remain in the
scoring denominator and receive zero.

The Evidence Auditor checks report structure, criterion coverage, evidence
availability, and evidence-type admissibility. A behavioral claim requires
execution-grounded evidence; source inspection suffices only for criteria
designated source-verifiable. Semantic verdicts remain the judge's
responsibility.

An audit finding identifies the affected task or criterion and the evidence
defect. The judge may perform up to three targeted remediation rounds for
affected criteria, retaining the initial output, audit findings, additional
evidence, and revised verdicts. A verdict changes only when new evidence
supports the revision. After the final round, a criterion or pairwise task
without a valid verdict or admissible evidence is assigned zero; accepted results for unrelated
criteria remain available. The scorer computes \textsc{ATS}, \textsc{CTPR},
and \textsc{ASR} from the audited results and fixed scoring policy.

\begin{figure}[H]
    \centering
    \includegraphics[width=\linewidth]{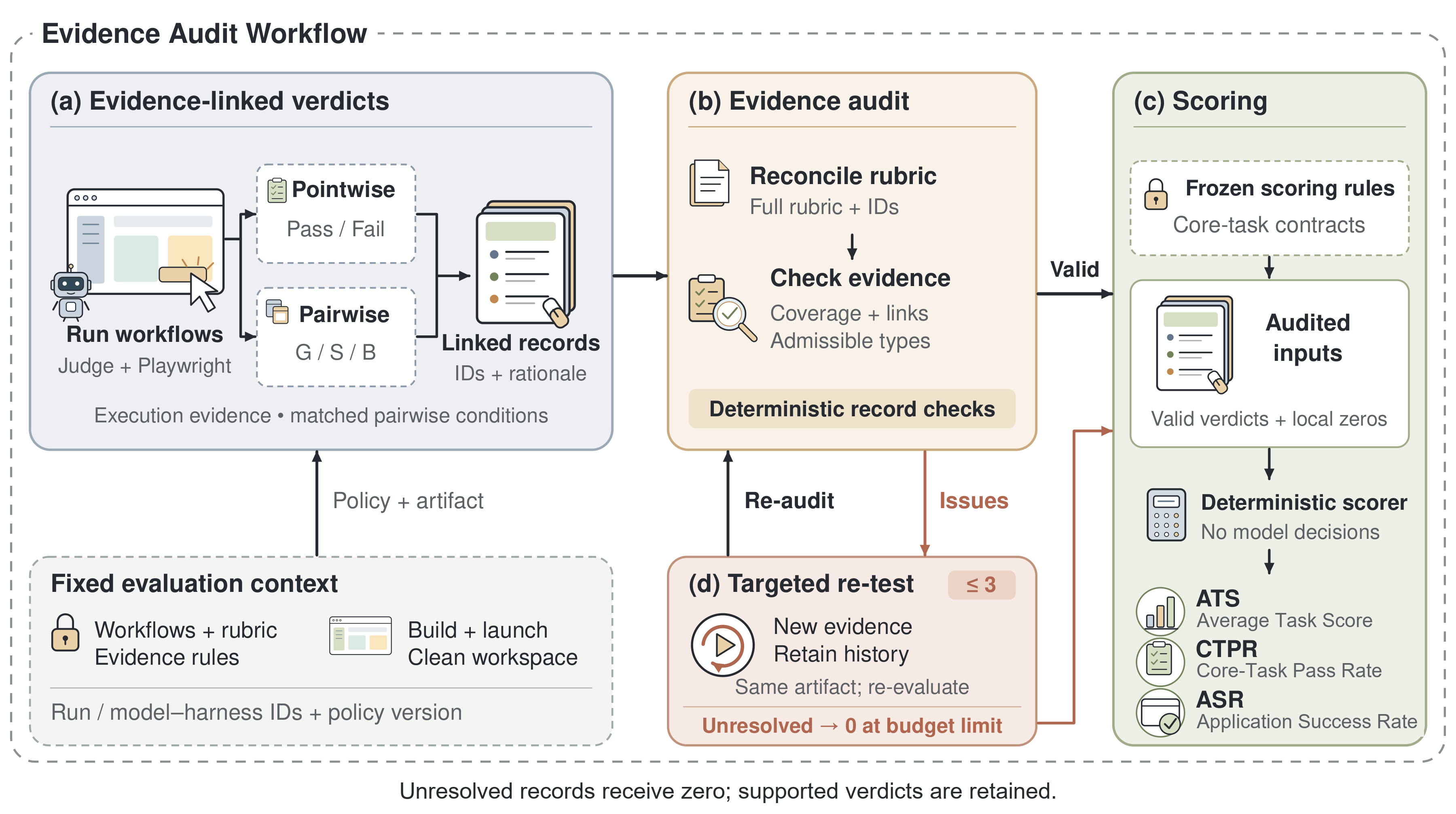}
    \caption{Evidence audit workflow from structured verdicts through up to
    three targeted remediation rounds to scoring. The limit applies to audit
    remediation rounds, not to individual test actions.}
    \label{fig:appendix-c-evidence-audit}
\end{figure}

\section{Metric Definitions and Threshold Sensitivity}
\label{app:metric-sensitivity}

\subsection{Metric Semantics and Formal Definitions}

\begin{table}[H]
\centering
\footnotesize
\caption{Evaluation scope, unit scoring, and task aggregation for ATS,
\textsc{CTPR}, and \textsc{ASR}.}
\label{tab:metric_semantics}
\begin{tabularx}{\linewidth}{@{}l>{\raggedright\arraybackslash}X>{\raggedright\arraybackslash}X>{\raggedright\arraybackslash}X>{\raggedright\arraybackslash}X@{}}
\toprule
Metric & Evaluated scope & Unit score & Across-task aggregation & Interpretation \\
\midrule
ATS           & All tasks     & Graded      & Average     & Local completion and quality \\
\textsc{CTPR} & Core pointwise tasks & Binary pass & Average     & Average core-task acceptance \\
\textsc{ASR}  & Operability and all core tasks & Binary pass & Conjunction & Application-level success \\
\bottomrule
\end{tabularx}
\end{table}

For a fixed model--harness condition, let $\mathcal{Q}$ be the reported problem
set and let $R=3$ be the number of scheduled runs per problem. Let
$\mathcal{T}_q^{\mathrm{pt}}$ and $\mathcal{T}_q^{\mathrm{pw}}$ denote the
pointwise and pairwise tasks of problem $q$, respectively, with
$\mathcal{T}_q=\mathcal{T}_q^{\mathrm{pt}}\cup
\mathcal{T}_q^{\mathrm{pw}}$.

\Needspace{6\baselineskip}
For pointwise task $t$, let
$\mathcal{J}_{qt}$ be its atomic criteria. Its completion score is
\[
A_{qrt}
=
\frac{1}{|\mathcal{J}_{qt}|}
\sum_{j\in\mathcal{J}_{qt}} y_{qrtj},
\]
where $y_{qrtj}=1$ if criterion $j$ retains a Pass verdict after auditing and
$0$ otherwise. Failed, missing,
malformed, errored, or inadmissible criterion outcomes are assigned zero. For a
pairwise task, define
\[
G_{qrt}
=
\begin{cases}
1, & \text{Good},\\
0.5, & \text{Same},\\
0, & \text{Bad}.
\end{cases}
\]
Invalid or missing pairwise outcomes are likewise assigned zero.

For either judgment mode, define the task score as
\[
Z_{qrt}
=
\begin{cases}
A_{qrt}, & t\in\mathcal{T}_q^{\mathrm{pt}},\\
G_{qrt}, & t\in\mathcal{T}_q^{\mathrm{pw}}.
\end{cases}
\]

The problem-run ATS is
\[
O_{qr}
=
\frac{1}{|\mathcal{T}_q|}
\sum_{t\in\mathcal{T}_q}Z_{qrt}.
\]

Let $\mathcal{C}_q\subseteq\mathcal{T}_q^{\mathrm{pt}}$ be the predefined core
tasks and let $\mathcal{M}_{qt}\subseteq\mathcal{J}_{qt}$ contain the mandatory
criteria of core task $t$. Its pass indicator at threshold $\tau$ is
\[
P_{qrt}^{(\tau)}
=
\left(\prod_{j\in\mathcal{M}_{qt}}y_{qrtj}\right)
\mathbb{I}[A_{qrt}\geq\tau],
\]
and the problem-run \textsc{CTPR} is
\[
C_{qr}^{(\tau)}
=
\frac{1}{|\mathcal{C}_q|}
\sum_{t\in\mathcal{C}_q}P_{qrt}^{(\tau)}.
\]

Let $B_{qr}\in\{0,1\}$ indicate whether the generated application satisfies
the build-and-runtime contract, with a rationale and admissible evidence
retained after auditing. The problem-run application-level success indicator is
\[
S_{qr}^{(\tau)}
=
B_{qr}\prod_{t\in\mathcal{C}_q}P_{qrt}^{(\tau)}.
\]

For any problem-run quantity $x_{qr}$, both splits first average all $R=3$
scheduled run slots within each problem and then macro-average problems:
\[
\operatorname{Avg}(x)
=
\frac{1}{|\mathcal{Q}|}
\sum_{q\in\mathcal{Q}}
\frac{1}{R}
\sum_{r=1}^{R}x_{qr}.
\]
The reported metrics are
\[
\mathrm{ATS}=\operatorname{Avg}(O),\qquad
\mathrm{CTPR}_{\tau}=\operatorname{Avg}(C^{(\tau)}),\qquad
\mathrm{ASR}_{\tau}=\operatorname{Avg}(S^{(\tau)}).
\]

Let $\mathcal{D}$ denote the eight capability dimensions, let $d_{qt}$ be the
dimension assigned to task $t$ of problem $q$, and define
\[
\mathcal{U}_d
=
\{(q,t):q\in\mathcal{Q},\ t\in\mathcal{T}_q,\ d_{qt}=d\}.
\]
The capability score first averages each task across its scheduled
runs and then gives equal weight to all tasks assigned to dimension $d$:
\[
\mathrm{CapScore}_d
=
\frac{1}{|\mathcal{U}_d|}
\sum_{(q,t)\in\mathcal{U}_d}
\frac{1}{R}
\sum_{r=1}^{R}Z_{qrt},
\qquad d\in\mathcal{D}.
\]
These eight task-weighted capability scores support diagnostic analysis;
the main ranking uses application success.

We use $\tau=0.8$ for the primary evaluation and abbreviate
$\mathrm{CTPR}_{0.8}$ and $\mathrm{ASR}_{0.8}$ as \textsc{CTPR} and
\textsc{ASR}; per-run quantities also omit the threshold superscript at this
setting. Infrastructure failures are rerun. A model-generated artifact
that fails to build or run remains in the evaluation matrix with $B_{qr}=0$.
All scheduled slots remain in the denominator. Unresolved missing runs or
records invalidated as a whole receive zero for ATS, \textsc{CTPR}, and \textsc{ASR}.
CTPR and ASR require a nonempty core-task set for each evaluated problem;
otherwise, these metrics are unavailable.

\subsection{Public-Set Results and Failure Composition}
\label{app:validation-metrics}

On the 30-problem public set, OpenHands yields five-model means of 86.90\%
ATS, 79.36\% \textsc{CTPR}, and 50.89\% \textsc{ASR}
(Table~\ref{tab:validation_metrics}). The 28.47-point
\textsc{CTPR}--\textsc{ASR} difference contrasts average core-task acceptance
with joint application acceptance, which also requires operability.

\begin{table}[H]
\centering
\small
\caption{Aggregate OpenHands scores (\%) on the 30-problem public set.
\textsc{CTPR}$-$\textsc{ASR} contrasts average core-task pass rate with
joint application success.}
\label{tab:validation_metrics}
\begin{tabular*}{\linewidth}{@{\extracolsep{\fill}}lrrrr@{}}
\toprule
Model & ATS & \textsc{CTPR} & \textsc{ASR} & \textsc{CTPR}$-$\textsc{ASR} \\
\midrule
\qwenmodelcell{}     & 88.57 & 82.19 & 57.78 & 24.41 \\
Kimi K3         & 87.87 & 81.52 & 55.56 & 25.96 \\
GPT-5.6 Sol     & 90.53 & 82.33 & 55.56 & 26.78 \\
\deepseekmodelcell{} & 84.99 & 79.06 & 46.67 & 32.39 \\
GLM-5.3         & 82.54 & 71.69 & 38.89 & 32.80 \\
\midrule
Mean            & 86.90 & 79.36 & 50.89 & 28.47 \\
\bottomrule
\end{tabular*}
\end{table}

Table~\ref{tab:failure_mechanisms} reports failure mechanisms among unsuccessful
OpenHands runs on the public set with valid criterion judgments. A mandatory
failure occurs when any core task fails a mandatory criterion. A completion
failure occurs when a core task passes all of its mandatory criteria but falls
below 80\% completion. ``Mandatory only'' and ``completion only'' indicate
one mechanism; ``both'' indicates a mandatory failure and a completion failure
in different core tasks. The final row pools valid failed runs across models.
The three categories cover the valid failed runs in this analysis.
Invalid reports are excluded from this conditional analysis and retain their
zero scores in the scheduled-run metrics.

\subsection{Representative Cases for the Three Metric Layers}
\label{app:metric-case-evidence}

Table~\ref{tab:metric-case-summary} gives the per-run scores behind
the two public-set runs discussed in Section~\ref{sec:application-failures}.
Both reports contain audited criterion judgments; each run fails its
core-task acceptance conditions and has $S_{qr}=0$.

\subsubsection{Case A: High Task Scores with Failed Mandatory Rules}
The specification requires a personal-finance dashboard that lets users
complete a transfer while keeping the balance, transaction list, and derived
statistics consistent.
The minimum workflow for that goal includes amount validation, secondary
confirmation, transaction submission and write-back, balance deduction, and
updates to the derived views. Two mandatory criteria require invalid amounts to be
rejected before confirmation and all derived views to reflect a completed
transfer. Each criterion is required for its core task to pass.

\begin{figure}[H]
\centering
\includegraphics[width=0.31\linewidth,trim=0 0 0 440,clip]{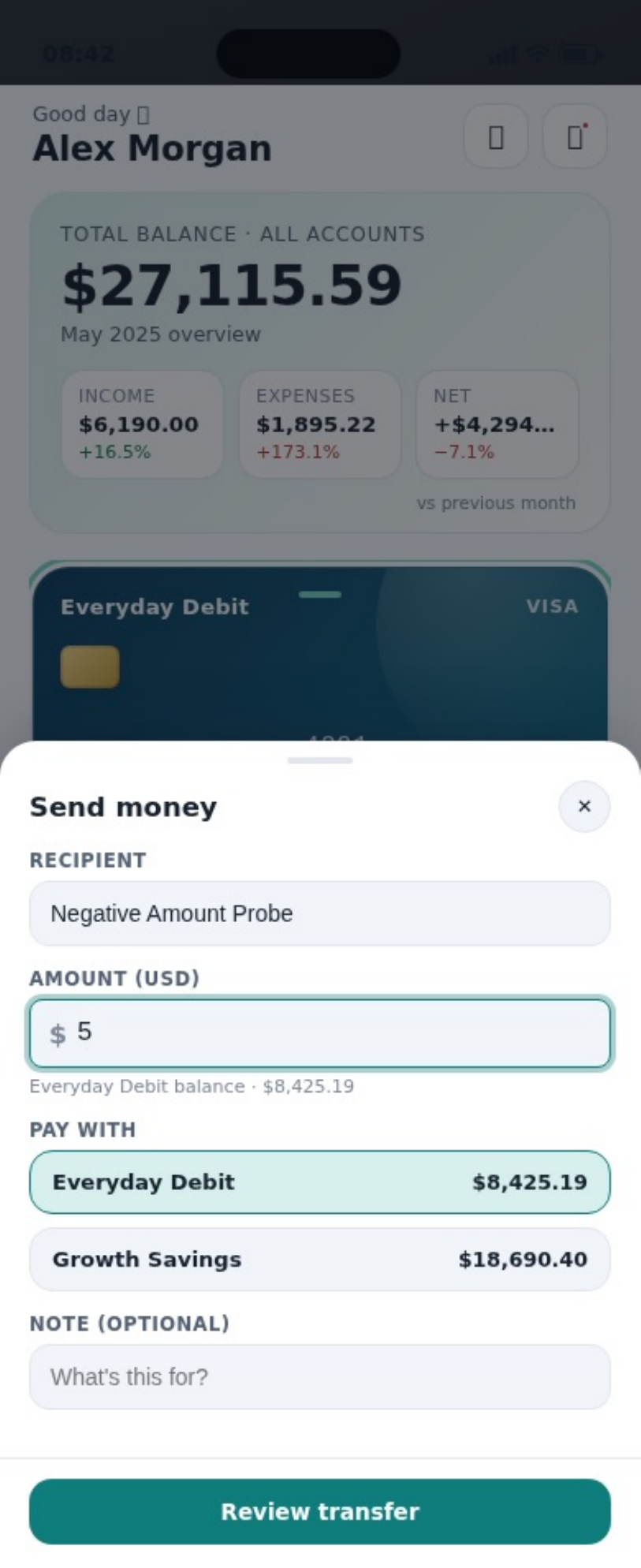}
\caption{Case A: after entering \texttt{-5}, only \texttt{5} remains in the
amount field with no validation error. Figure~\ref{fig:representative-failures}(a)
shows the resulting \$5.00 confirmation.}
\label{fig:case-a-negative}
\end{figure}

This run fails at both points. Entering \texttt{-5} silently removes the minus
sign and leaves \texttt{5} in the amount field without an error
(Figure~\ref{fig:case-a-negative}), then proceeds to a \$5.00 secondary
confirmation (Figure~\ref{fig:representative-failures}(a)). After a
valid transfer, the ``Transfer'' entry in the category breakdown rises from
\$120.00 over one transaction to \$132.34 over two. The transaction therefore
appears in some derived views, but the category budgets still have four rows
with no Transfer budget, and overall use remains at 49\%. The budget view does
not reflect the updated transaction state.

Across all seven tasks, five score 100 and two score 66.67, so
\textsc{ATS} is $(5\times100+2\times66.67)/7\approx90.48$. The two failures lower
\textsc{ATS} by only 9.52 points, but each failed mandatory criterion prevents
its core task from passing. Only two of four core tasks pass; thus
the per-run core-task pass fraction is $C_{qr}=50\%$ and $S_{qr}=0$.

\subsubsection{Case B: Passing Mandatory Criteria with Incomplete Core Tasks}
The problem asks for a declaratively animated octopus playing a jazz drum kit
in a single native-SVG file. Its core tasks require eight arms with distinct
playing roles and stick heads that reach their corresponding drums or cymbals.
Six mandatory criteria check whether this basic action sequence and its implementation constraints
hold. All six pass in this run.

\begin{figure}[H]
\centering
\includegraphics[width=0.67\linewidth]{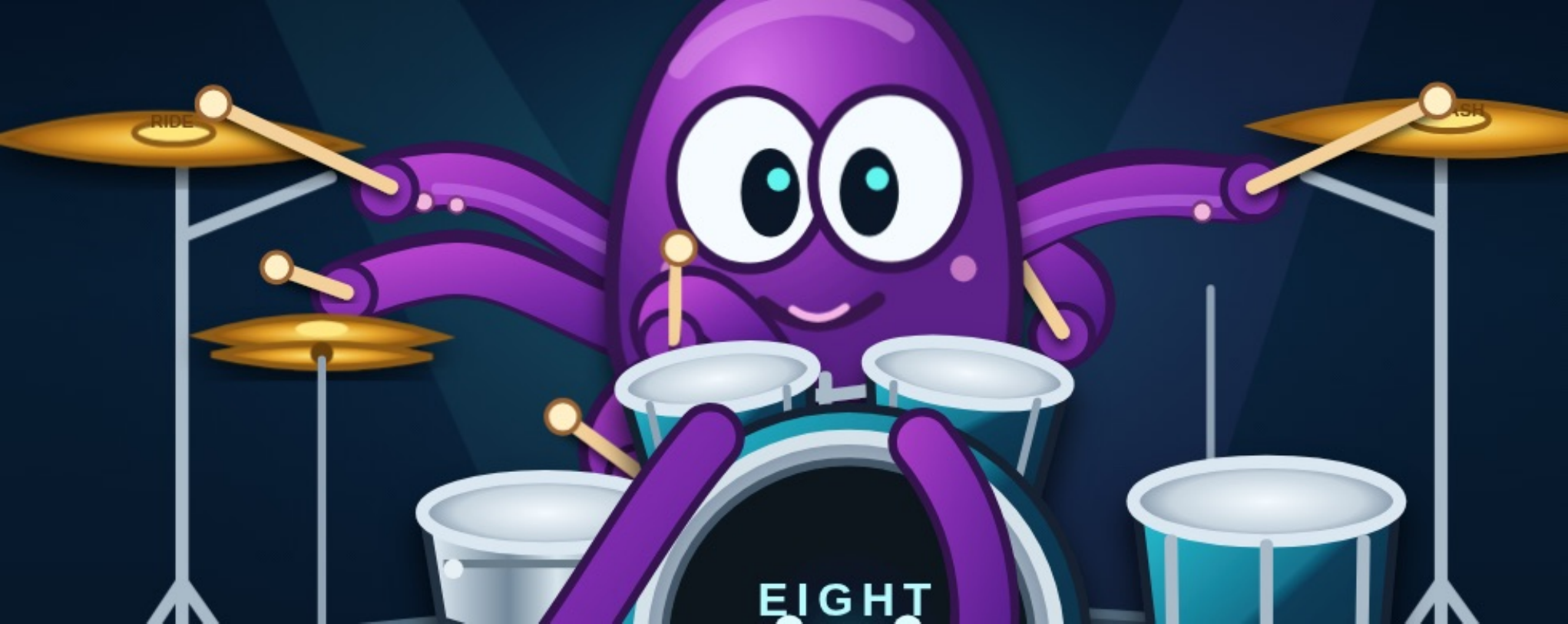}
\caption{Case B: the eight animated arms have equal-width paths and lack
natural taper (sampled at $t=0.12$). The ride contact frame at $t=2.34$ is
shown in Figure~\ref{fig:representative-failures}(b).}
\label{fig:metric-case-b}
\end{figure}

The core tasks also include non-mandatory details. Task 2 scores 75.00\%. The eight
arms have distinct roles and remain attached, but their equal-width paths lack
natural taper (Figure~\ref{fig:metric-case-b}). Task 4 scores 50.00\%. The sticks
reach the snare, rack tom, ride, and crash, but the struck surfaces do not deform or move
in synchrony (Figure~\ref{fig:representative-failures}(b)). In runtime samples, the
struck targets' bounding boxes are bit-identical across nine instants, while
the kick beater changes width in the same data. This confirms that the sampling
procedure captured motion elsewhere in the same artifact. The remaining
defects concern the specified arm shape and synchronized target response.

Although all six mandatory criteria pass, each core task must also reach
80\% completion. Tasks 2 and 4 score 75.00\% and 50.00\%, so only one of
three core tasks passes: $C_{qr}=33.33\%$ and $S_{qr}=0$.
Appendix~\ref{app:threshold-sensitivity} examines the 80\% threshold.

Case A violates mandatory transfer rules; Case B satisfies its mandatory
criteria but has incomplete core tasks. The cases illustrate the distinct
roles of mandatory criteria and the completion threshold in application
acceptance.

\FloatBarrier

\subsection{Completion-Threshold Sensitivity}
\label{app:threshold-sensitivity}
\label{sec:threshold-choice}

We compared $\tau\in\{0.75,0.80,0.90\}$ on the 30-problem public set under
OpenHands across the five evaluated models. Raising the threshold from 0.75 to
0.80 lowers the five-model mean \textsc{ASR} by 7.33 percentage points and
mean \textsc{CTPR} by 3.56 points. Raising it from 0.80 to 0.90 lowers them
by a further 6.00 and 3.01 points, respectively. All changes are computed
before rounding.

\begin{table}[H]
\centering
\small
\caption{Completion-threshold sensitivity on the 30-problem public set under
OpenHands. Means average five models; ASR range is the maximum minus minimum
model ASR, in percentage points. All scores are percentages.}
\label{tab:threshold_sensitivity}
\begin{tabular*}{\linewidth}{@{\extracolsep{\fill}}lrrr@{}}
\toprule
Threshold & Mean \textsc{ASR} & \textsc{ASR} range & Mean \textsc{CTPR} \\
\midrule
0.75 & 58.22 & 17.78 & 82.91 \\
0.80 & 50.89 & 18.89 & 79.36 \\
0.90 & 44.89 & 21.11 & 76.34 \\
\bottomrule
\end{tabular*}
\end{table}

The completion threshold operates discretely because each core task contains
a finite number of criteria. On the public set, a threshold of
0.90 requires all criteria to pass for 99.1\% of core tasks. At 0.80, this
proportion is 86.6\%, while the remaining 13.4\% permit one non-mandatory criterion
to fail. Table~\ref{tab:threshold_sensitivity} shows how the alternative
thresholds affect aggregate scores. Considering these score changes and the
discrete behavior of the criteria, we use 0.80 as the default completion
threshold.

\section{Additional Results and Diagnostics}
\label{app:full-results}

The additional test-set results cover application categories, capability
dimensions, and repeated-run outcomes across 70 problems. Generation-resource
statistics use the public set.

\subsection{ASR by Application Category}
\label{app:category-asr}
\label{sec:category-performance}

\begin{figure}[H]
\centering
\includegraphics[width=\linewidth]{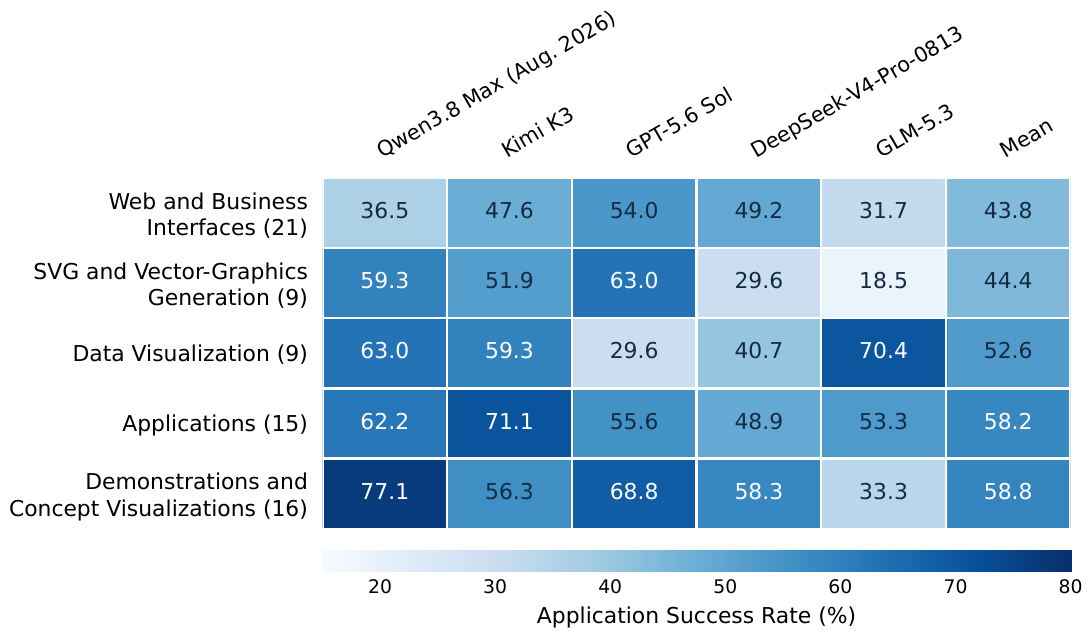}
\caption{Category-wise \textsc{ASR} (\%) on the 70-problem test set under
OpenHands. Cells show scores and cross-model means; row labels give problem
counts. Color represents \textsc{ASR}.}
\label{fig:category_performance}
\end{figure}

Figure~\ref{fig:category_performance} displays category-level \textsc{ASR},
problem counts, and cross-model means.

Across the evaluated test problems, Web and Business Interfaces and SVG and
Vector-Graphics Generation have the lowest cross-model means (43.8\% and
44.4\%).

\Needspace{27\baselineskip}
\subsection{Capability-Dimension Scores}
\label{app:capability-results}
\label{sec:capability-dimensions}

Table~\ref{tab:capability_scores} gives task counts, full capability scores, and
dimension means for Figure~\ref{fig:capability_scores_heatmap}.

\begin{table}[H]
\centering
\small
\caption{Capability scores (\%) on the 70-problem test set under
OpenHands. Parentheses give task counts; bold marks the highest value within
each dimension.}
\label{tab:capability_scores}
\setlength{\tabcolsep}{3pt}
\renewcommand{\arraystretch}{1.15}
\begin{tabular*}{\linewidth}{@{\extracolsep{\fill}}lrrrr@{}}
\toprule
Model & \multicolumn{1}{c}{\shortstack{Component\\Completeness\\(75)}} & \multicolumn{1}{c}{\shortstack{Representation\\Fidelity\\(30)}} & \multicolumn{1}{c}{\shortstack{Visual\\Quality\\(65)}} & \multicolumn{1}{c}{\shortstack{Interaction\\Responsiveness\\(27)}} \\
\midrule
\qwenmodelcell{} & 88.44 & 94.44 & 73.85 & \textbf{98.77} \\
Kimi K3 & \textbf{90.67} & 94.44 & 72.31 & 92.59 \\
GPT-5.6 Sol & 88.89 & \textbf{96.67} & \textbf{83.08} & 95.06 \\
\deepseekmodelcell{} & 86.22 & 93.33 & 59.49 & 96.30 \\
GLM-5.3 & 87.56 & 83.33 & 63.59 & 93.83 \\
\midrule
Mean & 88.36 & 92.44 & 70.46 & 95.31 \\
\bottomrule
\addlinespace[6pt]
\toprule
Model & \multicolumn{1}{c}{\shortstack{Computational\\Rendering\\(77)}} & \multicolumn{1}{c}{\shortstack{Rule\\Simulation\\(167)}} & \multicolumn{1}{c}{\shortstack{State\\Management\\(84)}} & \multicolumn{1}{c}{\shortstack{Data\\Exchange\\(20)}} \\
\midrule
\qwenmodelcell{} & \textbf{92.64} & 92.42 & 90.48 & \textbf{98.33} \\
Kimi K3 & 91.34 & 91.42 & \textbf{91.27} & 95.00 \\
GPT-5.6 Sol & 91.77 & \textbf{92.61} & 89.29 & 86.67 \\
\deepseekmodelcell{} & 88.74 & 88.82 & 90.87 & 91.67 \\
GLM-5.3 & 86.15 & 84.63 & 88.49 & 93.33 \\
\midrule
Mean & 90.13 & 89.98 & 90.08 & 93.00 \\
\bottomrule
\end{tabular*}
\end{table}

\subsection{Generation Resource Use}
\label{app:efficiency}

For runs on the 30-problem public set, mean generation latency is 39.65
minutes under OpenCode and 59.24 minutes under OpenHands; mean token use per
run is 6.18M and 9.52M, respectively. These describe generation resources
for harnesses with different agent loops (Appendix~\ref{app:harness-config}).

\subsection{Repeated-Run Outcome Distributions}
\label{app:reliability}

Table~\ref{tab:main_leaderboard} summarizes repeated-run outcomes on the
70-problem test set. Each of the five models has three scored OpenHands
generation-and-evaluation runs for every test problem, giving 70 model--problem
conditions and 210 scored runs per model. Each condition is categorized as all
successful, mixed, or all failed. These distributions describe end-to-end
outcome consistency across three runs, while \textsc{ASR} summarizes per-run
success under the scoring protocol in Appendix~\ref{app:metric-sensitivity}.

\subsection{Additional End-to-End Failure Cases}
\label{app:end-to-end-cases}

The following public-set cases supplement the detailed metric examples
in Appendix~\ref{app:metric-case-evidence}. In Q2, restoring a field to its
template value should also remove the override and restore inheritance; 13
of 15 OpenHands reports record a failure on this criterion. The visible value
can be correct while its source remains wrong. In Q20, 13 of 15 reports record
synchronization failures after a transaction: the list and balance may update without matching changes
to charts, budgets, timelines, or aggregate statistics. Both cases expose
inconsistent state semantics despite working local interactions.

Across the evaluated artifacts for Q12, Q18, and Q29, observed failures involve
contact, phase, direction, or dependence on velocity and gravity. Examples include
timeline-driven vapor-film effects, projectile arcs whose landing points are
disconnected from input parameters, and moving drumsticks without actual
contact with their targets. These failures concern the specified mechanisms
and their consistency over time.

\section{Agreement with the Adjudicated Human Reference}
\label{app:human-materials}

\subsection{Comparison Scope}
\label{sec:human_method}

The analysis covers all 30 public-set problems and 726 pointwise
atomic criteria. The pointwise track compares Pass/Fail judgments, while the pairwise track
compares Good/Same/Bad judgments on 15 tasks. Score-level comparisons use
aggregated pointwise judgments. The comparison measures agreement in criterion
verdicts and pointwise scores; application-level human agreement is a separate
measurement target.

\paragraph{Human reference.}
Two evaluators independently annotate each unit under a shared protocol,
without access to the identity of the generating model or one another's
decisions. Agreed labels become the reference directly; a third evaluator
adjudicates disagreements. The automated judge completes its assessment
without access to human labels or rationales. Automated outputs are matched
to the final human reference after both assessments are complete.

\subsection{Agreement Results}
\label{sec:alignment_results}

Table~\ref{tab:human_agreement} consolidates the reported label- and score-level
comparisons.

\begin{table}[H]
\centering
\small
\caption{Agreement between the automated judge and the adjudicated human reference on 30
public-set problems. Pointwise uses Pass/Fail atomic criteria;
pairwise uses Good/Same/Bad judgments on 15 tasks. Problem- and task-level
scores aggregate pointwise judgments. Pooled micro agreement combines the two
label tracks and is supplementary. Formulas appear in
Appendix~\ref{app:alignment-formulas}.}
\label{tab:human_agreement}
\begin{tabular*}{\linewidth}{@{\extracolsep{\fill}}llr@{}}
\toprule
Metric & Evaluation unit & Value \\
\midrule
\multicolumn{3}{@{}l}{\textbf{Label-level agreement}} \\
Exact agreement       & Pointwise criteria & 83.47\% \\
Cohen's $\kappa$       & Pointwise criteria & 0.651 \\
Macro-F1              & Pointwise criteria & 0.825 \\
Exact agreement       & Pairwise tasks ($n=15$) & 86.67\% \\
Micro exact agreement & Pooled tracks      & 83.54\% \\
\midrule
\multicolumn{3}{@{}l}{\textbf{Score association and error}} \\
Pearson $r$           & Problem           & 0.807 \\
Spearman $\rho$       & Problem           & 0.766 \\
RMSE/SD               & Problem           & 0.631 \\
Mean error            & Problem           & $+0.0362$ \\
Pearson $r$           & Task               & 0.709 \\
Spearman $\rho$       & Task               & 0.731 \\
\bottomrule
\end{tabular*}
\end{table}

\subsection{Metric Formulas}
\label{app:alignment-formulas}

\begingroup
\small
\setlength{\abovedisplayskip}{4pt}
\setlength{\belowdisplayskip}{4pt}
\setlength{\abovedisplayshortskip}{3pt}
\setlength{\belowdisplayshortskip}{3pt}

Label-level metrics in Section~\ref{sec:human-alignment} compare pointwise
Pass/Fail criteria or pairwise Good/Same/Bad tasks as separate tracks.
Score-level metrics aggregate pointwise judgments, with $x_i$ the final human
reference score and $y_i$ the judge score on the same problem or task; $n$
denotes the number of comparable units.

\paragraph{Exact agreement.}
\begin{equation}
\operatorname{Agreement}
=\frac{N_{\mathrm{match}}}{N_{\mathrm{comparable}}},
\label{eq:alignment_agreement}
\end{equation}
where $N_{\mathrm{match}}$ is the number of units assigned exactly the same
label by both sides and $N_{\mathrm{comparable}}$ is the number of units for
which both sides produced a valid label. This metric compares final labels;
$\kappa$ additionally accounts for agreement expected by chance.

\paragraph{Cohen's $\kappa$.}
\begin{equation}
\kappa
=\frac{p_o-p_e}{1-p_e},
\label{eq:pointwise_kappa}
\end{equation}
where $p_o$ is the observed agreement (i.e.\ exact agreement) and $p_e$ is the
chance agreement expected from the two Pass/Fail marginal distributions.
$\kappa=0$ indicates that observed agreement equals the chance agreement
expected from the marginal label distributions; $\kappa=1$ indicates perfect
agreement. We compute $\kappa$ for pointwise Pass/Fail judgments.

\paragraph{Macro-F1.}
\begin{equation}
\operatorname{\text{Macro-F1}}
=\frac{1}{2}\sum_{c\in\{\mathrm{Pass},\mathrm{Fail}\}}
\frac{2\,TP_c}{2\,TP_c+FP_c+FN_c}.
\label{eq:macro_f1}
\end{equation}
With adjudicated human labels as the reference and judge labels as predictions,
$TP_c$, $FP_c$, and $FN_c$ are the true positives, false positives,
and false negatives obtained when class $c$ is taken as the positive class.
Macro-F1 weights the Pass and Fail classes equally.

\paragraph{Pearson correlation.}
\begin{equation}
r
=\frac{\sum_{i=1}^{n}(x_i-\bar{x})(y_i-\bar{y})}
{\sqrt{\sum_{i=1}^{n}(x_i-\bar{x})^2
\sum_{i=1}^{n}(y_i-\bar{y})^2}},
\label{eq:problem_pearson}
\end{equation}
where $\bar{x}=\frac{1}{n}\sum_{i}x_i$ and $\bar{y}=\frac{1}{n}\sum_{i}y_i$ are
the two score means. Pearson $r$ measures linear association and is invariant
to additive shifts and positive rescaling; RMSE/SD and mean error quantify
score discrepancies.

\paragraph{Spearman correlation.}
\begin{equation}
\rho
=\frac{\sum_{i=1}^{n}(R_i-\bar{R})(S_i-\bar{S})}
{\sqrt{\sum_{i=1}^{n}(R_i-\bar{R})^2
\sum_{i=1}^{n}(S_i-\bar{S})^2}},
\label{eq:problem_spearman}
\end{equation}
where $R_i=\operatorname{rank}(x_i)$ and $S_i=\operatorname{rank}(y_i)$ are the
ranks of the two score series and $\bar{R}$, $\bar{S}$ are their means.
Spearman $\rho$ measures monotonic association through the ordering of scores.

\paragraph{Normalized root-mean-square error.}
\begin{equation}
\frac{\operatorname{RMSE}}{\operatorname{SD}(x)}
=
\frac{\sqrt{\frac{1}{n}\sum_{i=1}^{n}(y_i-x_i)^2}}
{\sqrt{\frac{1}{n-1}\sum_{i=1}^{n}(x_i-\bar{x})^2}},
\label{eq:problem_rmse_sd}
\end{equation}
The numerator is the root-mean-square error of the judge scores against the
human scores; the denominator is the sample standard deviation of the human scores. The ratio
expresses score error relative to variation in the human reference. A value
below 1 indicates that RMSE is smaller than this standard deviation; a value
near 1 indicates comparable magnitudes.

\paragraph{Mean error (ME).}
\begin{equation}
\operatorname{ME}
=\frac{1}{n}\sum_{i=1}^{n}(y_i-x_i).
\label{eq:problem_me}
\end{equation}
$\mathrm{ME}>0$ indicates that judge scores are higher on average, and
$\mathrm{ME}<0$ indicates lower scores. ME is the signed average error;
positive and negative discrepancies may cancel.

\endgroup

\end{document}